\documentclass[aps,pra,prabib,twocolumn,showpacs,nofootinbib]{revtex4}
\usepackage{graphicx} \usepackage{amsmath} \usepackage{amssymb}
\usepackage{amsfonts} \usepackage{bm}

\begin{document}

\newcommand{\be}{\begin{equation}} \newcommand{\ee}{\end{equation}}
\newcommand{\bea}{\begin{eqnarray}}\newcommand{\eea}{\end{eqnarray}}


\title{Atom capture by nanotube and scaling anomaly}

\author{Pulak Ranjan Giri} \email{pulakranjan.giri@saha.ac.in}

\affiliation{Theory Division, Saha Institute of Nuclear Physics,
1/AF Bidhannagar, Calcutta 700064, India}

\begin{abstract}
The existence of bound state of the polarizable neutral atom in the
inverse square potential created by  the electric field of a single
walled charged carbon nanotube (SWNT) is shown to be theoretically
possible. The consideration of inequivalent boundary conditions due
to self-adjoint extensions  lead to this nontrivial bound state
solution. It is also shown that the scaling anomaly is responsible
for the existence of such bound state. Binding of the polarizable
atoms in the coupling constant interval $\eta^2\in[0,1)$ may be
responsible for the smearing of the edge of  steps in quantized
conductance, which has not been   considered so far in  the
literature.
\end{abstract}



\date{\today}

\maketitle

Research activities in atomic and molecular physics \cite{try,den1}
is speeding up due to the advancement of laser cooling technique
\cite{arimondo,arimondo1,arimondo2,arimondo3}. Substantial amount of
people are engaged in devising suitable confining mechanism
\cite{lene,lene1,lene2,lene3} for cold atoms and molecules so that
it can be stored at a  temperature down to nano-Kelvin. Among
different schemes for trapping and storing cold atoms; magnetic
trap, dipole trap, optical trap etc are important. Present day
technology is well equipped to handle atoms and molecules at
nano-Kelvin temperature. Study of different properties of these
super cold neutral  atoms or molecules in electric or magnetic
fields, created by charged wire \cite{den,gong}, current carrying
conductor or by some other mechanism is now feasible.

Quantized conductance(QC)  \cite{try} is one  such important
property, which is usually seen when atoms are moving in the
neighborhood of a charged nanotube. QC is basically discrete quantum
steps in the cross section for atom capture versus voltage of the
thin wire. Since the steps result from the angular momentum
quantization in the attractive  inverse square potential experienced
by the atom, it is also called ``angular momentum quantum ladder".
The edge of the steps are  not  infinitely sharp but a little bit
smeared out \cite{try}. This exotic behavior is usually attributed
to the tunneling of the neutral  atoms through the inverse square
potential $(\eta^2-1/4)/r^2$ and happens near the  value $\eta=0$ of
the coupling constant.  This conclusion is based on the usual
boundary condition that the wave-function is zero at the
singularity.  But we need further quantum mechanical  investigation,
because the peculiar nature of the inverse square potential in the
interval $\eta\in[0,1)$ may give rise to bound state due to a
nontrivial boundary condition which is known for a long time in
mathematical physics.

Quantum mechanical behavior of the inverse square potential is
subtle \cite{landau} in the sense that  it lies between $1/r^{n>2}$
and $1/r^{n<2}$. It should be noted that for $n>2$ usually there can
not be any bound states and for $n<2$ the potential is capable of
forming  bound states. Usually a particle moving in inverse square
potential does not form bound state or more specifically the system
has negative infinite ground state in the region $\eta^2< 0$
\cite{case,case1}. It is however possible to form a single bound
state for $\eta^2\in[0,1)$ by considering nontrivial boundary
condition at the singularity of the potential. In this context it
should be noted that the smearing of the edge of steps in QC also
occurs near $\eta=0$. This is a possible hint that the the effect of
smearing may have its origin in the nontrivial nature of the
boundary condition.

Therefore the subject of this paper is to study the behavior of the
neutral atom  in $(\eta^2-1/4)/r^2$ potential created by the charged
single-walled carbon nano-tube (SWNT) and investigate the possible
origin for the smearing in QC. However our analysis is equally
applicable to an atom or a  molecule moving in a current carrying
wire or in ferromagnetic wire \cite{tka}, where the magnetic
interaction between the atom or molecule and the wire is responsible
for the $(\eta^2-1/4)/r^2$ potential. It should be noted that in reality the
actual interaction is  highly nonlinear  due to multi-pole
expansion of the charge distribution of the polarized atom in the
SWNT field. But  the dominat part of the interaction is inverse square 
in nature.  The
other interactions are  short range interaction between the
atom and SWNT. The inverse square
interaction  is therefore responsible for the description of  
the long distance behavior of
the atom.  It is known
\cite{kumar1,kumar2,kumar3} that inverse square inreraction may form bound
state. Short distance interactions can however
be taken into account through  the boundary conditions imposed on
the Hamiltonian, keeping  self-adjointness \cite{reed,reed1} of the
Hamiltonian intact. Our goal is to find out suitable boundary
conditions for the Hamiltonian in the inverse square interaction
potential by using von Neumann's theory \cite{reed,reed1}  of
self-adjoint extensions, so that that the system remains
self-adjoint. Inverse square interaction has diverse applications
starting from microscopic physics to black holes
\cite{biru,kumar1,kumar2,falomir,falomir1,falomir2,kumar3,feher,bh,bh1,stjep}.
In our present work we use this model to explain the polarizable
neutral atom capture by charged nanotube(SWNT) and thereby the
possible occurrence of smearing in QC. von Neumann's method of
self-adjoint extensions suggest the possibility of the existence of
a single bound state. The
coefficient of the inverse square interaction is dimensionless. The
consequences of the  absence of the dimensionfull coefficients  in
the Hamiltonian imply that there should not be any bound state in
the system. But due to the quantum mechanical scaling anomaly
\cite{camblong1,camblong2,camblong3}, it can be shown that the
system can form bound states. The consideration of inequivalent
boundary conditions (self-adjoint extensions) is responsible  for
quantum mechanical anomalies \cite{esteve}, which in this case
allows the charged nanotube (SWNT) to capture polarizable neutral
atom. 

We consider a polarizable neutral atom of mass $\mu$ and
polarizability $\alpha$ moving in the electric field of a charged
single-walled carbon nanotube with line charge $q$. The electric
field $E$ of the charged nanotube induces a dipole moment $d= \alpha
E$ on the neutral atom. The interaction between this induced dipole
moment $d$ of the atom and the electric field of the nanotube
generates a potential $V_d= -\frac{1}{2}\alpha E^2=-2\alpha q^2
\frac{1}{r^2}$. For the cylindrical symmetry of the system, it is
convenient to take the $z$ axis along the nanotube. In  cylindrical
polar coordinates $(r,\phi, z)$, one can  write the Schr\"{o}dinger
equation for the system as follows:
\begin{equation}
\left[-\frac{\hbar^2}{2\mu}\nabla^2 - \frac{2\alpha
q^2}{r^2}\right]\Psi= \mathcal E\Psi\,, \label{schrodinger}
\end{equation}
\begin{figure}
\includegraphics[width=0.4\textwidth, height=0.2\textheight]{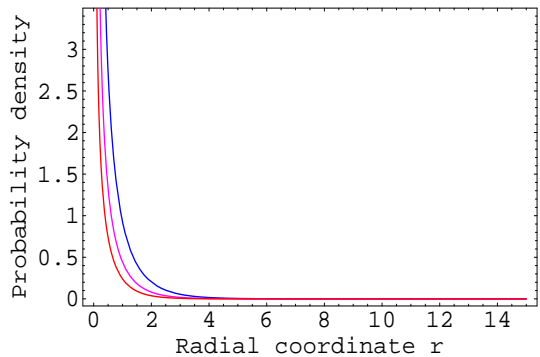}
\caption {(color online) A plot of the probability density of bound
state as a function of the radial coordinate $r$. For blue graph
$\omega=\pi/5, \alpha=1,
  q=\sqrt{1/25}, m= 1$. For pink graph $\omega=\pi/5, \alpha=3, q=\sqrt{1/25},
  m= 1$. For orange graph $\omega=\pi/5, \alpha=5, q=\sqrt{1/25}, m= 1$.}
\end{figure}

where $\mathcal E$ is the energy of the atom in the nanotube field.
We use the generic ansatz   $\Psi(r,\phi,z) = \frac{1}{\sqrt
r}R(r)e^{im\phi}e^{ikz}$, for the wave-function to separate the
Schr\"{o}dinger equation (\ref{schrodinger}).  The radial equation
after separating the angular part is \be H_r R(r) \equiv \left [
-\frac{d^2}{dr^2}+\frac{1}{r^2}\left(\eta ^2-\frac{1}{4}\right)
\right ]R(r) = \epsilon R(r)\,. \label{radial} \ee The radial
Hamiltonian $H_r$ has eigenvalue $\epsilon = 2\mu\mathcal
E/\hbar^2$.  $H_r$ depends on the coupling constant
$\eta^2=m^2-\lambda$, where $\lambda = 4\mu \alpha q^2/\hbar^2$. It
should be noted that in literature it is usually assumed that the
atom does not have any bound state in the the electric field of the
charged nanotube and depending on the sign of the effective coupling
constant $\eta^2-1/4$ the atom either falls to the center
\cite{landau} or escapes away from the nanotube towards the
infinity. However as pointed out earlier we show that  von Neumann's
technique allows us to make the Hamiltonian self-adjoint for
$-1/4\leq\eta^2-1/4<3/4\equiv \eta^2\in[0,1)$ and get a single bound
state for this self-adjoint Hamiltonian.

It is also to be noted that the above interval for the coupling
constant $\eta$ is actually realizable in experiment. Let us discuss
it with an example. Lithium atom can be a good candidate for our
example because it has been used in the experiment for cold atom
\cite{den1}. For the Lithium atom with mass $\mu_{Li}= 9.99\times
10^{-27}$Kg and polarization $\alpha_{Li}= 24\times
10^{-30}$$\mbox{m}^3$ \cite{den1} and a typical line charge density
$q= 5\times 10^{-8}$C/m, the coupling constant for angular momentum
quantum number $m=1$ is found to be $\eta_{Li}^2\approx 0.78\in
[0,1)$. It is also possible to get other values of $\eta_{Li}$, such
that it lies within the desired interval $\eta_{Li}^2\in[0,1)$ by
increasing the line charge density $q$. Note  that for  $m=0$, the
coupling constant $\eta_{Li}^2<0$ for any non-zero values of the
line charge density $q$. In this case it is already known
\cite{feher} (see section 6) that the spectrum is not bounded from
below.  Therefore the atom with $m=0$ will hit the nanotube  and
will  be lost.

Let us now come back to our discussion. In our case we are dealing
with a real symmetric (Hermitian) Hamiltonian $H_r$ over the domain
\begin{eqnarray}
\nonumber D(H_r) &\equiv & \\\{\phi (0) &=& \phi^{\prime} (0) = 0,~
\phi,~ \phi^{\prime}~ {\rm absolutely~ continuous} \}\,. \label{d1}
\end{eqnarray}
Since the domain (\ref{d1}) is too restricted, it fails to be
self-adjoint. Here we need to know that  if the Hamiltonian is
self-adjoint in a given domain $D$, then it should be equal to its
adjoint Hamiltonian domain $D^*$, i.e., $D=D^*$. In our case the
domain for our adjoint Hamiltonian $H^*_r$  is given by
\begin{eqnarray}
D(H^*_r) \equiv \{\phi~ |~ \phi,~ \phi^{\prime}~ {\rm absolutely~
continuous} \}\,. \label{d2}
\end{eqnarray}
 It is clear from the above expression of $D(H_r)$ and $D(H^*_r)$ that
$D(H_r)\neq D(H^*_r)$. So our Hamiltonian $H_r$ is not self-adjoint.
For a non self-adjoint operator, there exists complex eigenvalue
solutions for its adjoint operator, whereas  for a self-adjoint
operator there does not exist any complex eigenvalue solutions for
its adjoint operator \cite{reed,reed1}. So alternatively we can
determine whether $H_r $ is self-adjoint in $D(H_r)$, by looking  at
the square integrable solutions of the equations \be H_r^*
\phi_{\pm} = \pm i \phi_{\pm}, \label{def} \ee The adjoint operator
$H_r^*$ in our case is given by the same differential expression as
$H_r$,  but their domains are different. The square-integrable
solutions for the $+$ type and $-$ type of (\ref{def}) are called
the deficiency space solutions and the number of solutions $n_\pm$
are called deficiency indices. According to the numbers $n_+$ and
$n_-$, the Hamiltonian $H_r$ can be divided   into three classes:\\  
Self-adjoint(essentially):  for  $( n_+ , n_- ) = (0,0)$.\\ Admit self-adjoint
extensions: for $n_+ = n_- \neq 0$.\\ No self-adjoint extensions possible:  
for $n_+ \neq n_-$.\\
For the Hamiltonian $H_r$ we need to concentrate only on  
square-integrable  solutions of
(\ref{def}) for our analysis. The two solutions  which are square
integrable at infinity are
\begin{eqnarray}
\phi_+ (r) = r^{1/2}H^{(1)}_\eta (re^{i \frac{ \pi}{4}}), \phi_- (r)
= r^{1/2}H^{(2)}_\eta (re^{-i \frac{ \pi}{4}})\,. \label{hankel}
\end{eqnarray}
$H_\eta$'s in (\ref{hankel}) are Hankel functions \cite{abr}.  We
first consider the case $ \eta \neq 0$. The short distance behavior
($r \rightarrow 0$) of the functions $\phi_\pm$ are given by
\begin{eqnarray}
\nonumber \phi_+(r) \simeq
\mathcal{C}_1(\eta)\left(\frac{r}{2}\right)^{\eta+1/2} +
\mathcal{C}_2(\eta)\left(\frac{r}{2}\right)^{-\eta+1/2 },\\
\phi_-(r) \simeq \mathcal{C}^*_1(\eta)\left(\frac{r}{2}\right)^{\eta
+1/2} + \mathcal{C}^*_2(\eta)\left(\frac{r}{2}\right)^{-\eta+1/2},
\end{eqnarray}
where $\mathcal{C}_1(\eta)=\frac{i}{\sin\eta \pi}\frac{ e^{ - i
\frac{3 \eta
 \pi}{4}}}{\Gamma (1 + \eta)}$, $\mathcal{C}_2(\eta)=-\frac{i}{\sin\eta
 \pi}\frac{ e^{ - i \frac{\eta \pi}{4}}}{\Gamma (1 -\eta)}$ and
 $\mathcal{C}_1^*(\eta)$ and $\mathcal{C}_2^*(\eta)$ are complex conjugates of
 $\mathcal{C}_1(\eta)$ and $\mathcal{C}_2(\eta)$ respectively. The solutions
 $\phi_{\pm}$ are not square integrable near the singularity ($r\to 0$) for
 ${\eta}^2 \geq 1$.  This means the deficiency indices are zero, i.e.,  $n_+ =
 n_- = 0$.  So according to our classification scheme above, $H_r$ is
 essentially self-adjoint in the domain $D(H_r)$ \cite{reed,reed1} for ${\eta}^2
 \geq 1$.  But for $ -1 < \eta < 0$ or $ 0 < \eta < 1$, the deficiency space
 solutions $\phi_\pm$ are square-integrable. Which means, $H_r$ has deficiency
 indices $ (1,1)$ in this range and is not self-adjoint on the domain
 $D(H_r)$. Since the deficiency indices $n_+=n_-=1$ are equal, the Hamiltonian
 admits self-adjoint extensions. Since the case of zero deficiency indices
 i.e., $(0,0)$, indicates that the corresponding Hamiltonian is self-adjoint,
 any departure from $(0,0)$ shows the measure by which the  Hamiltonian
fails to
 be self-adjoint. In order to make it self-adjoint we then need to add the
 deficiency space solutions with the original domain and make it less
 restrictive such that the new domain becomes equal to
its corresponding adjoint domain. In our present case also since the
Hamiltonian $H_r$ is not
 self-adjoint in the domain $D(H_r)$, we need to add the deficiency space
 solutions $\phi_\pm$ to the domain $D(H_r)$ appropriately to make it less
 restrictive, such that the Hamiltonian in this new domain becomes
 self-adjoint. Thus the new domain $D_\omega(H_r)$ in which $H_r$ is
 self-adjoint is given by
 \begin{eqnarray}
 D_\omega(H_r)\equiv D(H_r) + \phi_+ + {\mathrm e}^{i\omega} \phi_-\,,
 \label{ndomain}
 \end{eqnarray}
 where $ \omega \in R$ (mod $2 \pi$) \cite{reed,reed1}. One can now evaluate the
 spectrum of the Hamiltonian $H_r$ taking the self-adjoint domain
 $D_\omega(H_r)$ into account.

One can evaluate the solution of the differential equation
(\ref{radial})  as
\begin{equation}
R(r)\equiv r^{1/2}H^{(1)}_\eta(\sqrt{\epsilon}r)\,, \label{radial1}
\end{equation}
The short distance  ($r \rightarrow 0$) behavior of $\phi_{+}(r) +
e^{i \omega} \phi_{-}(r)$ is given by
\begin{eqnarray}
\nonumber \phi_{+}(r) + e^{i \omega} \phi_{-}(r) \simeq \hspace{3cm}\\
\nonumber \frac{2e^{i\omega/2}}{\Gamma(1+\eta)}\frac{\sin(\omega/2+3
\pi\eta/4)}{\sin\pi\eta}\left(\frac{r}{2}\right)^{\eta+1/2}\\
  -
 \frac{2e^{i\omega/2}}{\Gamma(1-\eta)}\frac{\sin(\omega/2+\pi\eta/4)}
 {\sin\pi\eta}\left(\frac{r}{2}\right)^{-\eta+1/2}
\label{match1}
\end{eqnarray}
The short distance behavior of (\ref{radial1}) on the other hand is
given by
\begin{eqnarray}
\nonumber R(r) \simeq\hspace{5cm}\\
\mathcal{D}_1(\eta,\sqrt\epsilon)\left(\frac{r}{2}\right)^{\eta+1/2
} +
\mathcal{D}_2(\eta,\sqrt\epsilon)\left(\frac{r}{2}\right)^{-\eta+1/2},
\label{match2}
\end{eqnarray}
where $\mathcal{D}_1(\eta,\sqrt\epsilon)=
\frac{i}{\sin\pi\eta}\frac{e^{-i\pi\eta}{\sqrt{\epsilon}}^\eta}{\Gamma(1+\eta)}$
and $\mathcal{D}_2(\eta,\sqrt\epsilon)=
-\frac{i}{\sin\pi\eta}\frac{{\sqrt\epsilon}^{-\eta}}{\Gamma(1-\eta)}$.
In order that $R(r)$ belongs to $D_\omega(H_r)$,  the coefficients
of $r^{\eta+1/2 }$ and $r^{- \eta+1/2}$ in (\ref{match1}) and
(\ref{match2}) should match. The bound state energy can be found by
comparing the coefficients and the energy eigenvalue is given by
\begin{equation}
 E= - \frac{\hbar^2}{2\mu} \left [ \cos\frac{\pi\eta}{2}+
    \cot(\frac{\omega}{2}+\frac{\pi\eta}{4})\sin\frac{\pi\eta}{2}\right ]
    ^{\frac{1}{\eta}}\,. \label{eigen1}
\end{equation}
 From (\ref{eigen1}), we can see that the system given by the Hamiltonian
$H_r$ admits a single bound state solution. 
Note that not all values of $\omega$
allows a bound state. In fact only those values of $\omega$ admits a
bound state for which the quantity in first bracket in
(\ref{eigen1}) is positive. Since $\epsilon$ is negative, we can
write down the the bound state eigenfunction (\ref{radial1}) as
\begin{equation} R(r) \equiv r^{1/2}H^{(1)}_\eta(i \sqrt{|\epsilon|} r),
\end{equation}
The parameter $\omega$, in bound state eigenvalue and in the bound
state eigenfunction indicates that each value of parameter $\omega$
characterizes one separate system, thus leading to inequivalent
quantization.

One can also recalculate all the above things for $\eta = 0$. The
procedure is same. Here we only state the results for completeness.
The bound state energy and the wave function for $\eta=0$ are given
by
\begin{eqnarray}\nonumber E =
-\frac{\hbar^2}{2\mu} {\exp}\left[\frac{\pi}{2} {\cot}
\frac{\omega}{2}\right]\,,\\ R(r) = \sqrt{-2 \epsilon r} K_0\left(
\sqrt{-\epsilon} r\right)\,,
\end{eqnarray}
respectively, where $K_0$ is the modified Bessel function
\cite{abr}. Note that the above analysis shows that for $-1 < \eta <
1$, the radial Hamiltonian describing an atom in the electric field
$E$ of a charged nanotube admits a single bound state. From the
condition $-1<\eta<1$, we can easily check that the particle with
arbitrarily small polarizability can form bound state in the
electric field of charged nanotube provided the line charge is sufficient to
satisfy the condition $\eta^2\in[0,1)$. This conclusion is remarkably
different from the statement in the literature that either the atom
will fall into the charged nanotube or fly away depending upon the
sign of the  coupling constant $\eta^2$.  
Our model predicts the existence of a single
bound state. The exact numerical value of the bound state energy
would depend on the choice of the self-adjoint extension parameter
$\omega$ which characterizes the boundary conditions at the origin.
The bound state probability density for the atom has been plotted in
FIG.1. It shows the nontrivial nature of the boundary condition,
that the wave function is not zero at the origin, but still it is
square integrable. In FIG. 2,  bound state eigenvalue of the atom
has been plotted as a function of the polarizability of the atom. It
can be easily shown that the bound state eigenvalue of the atom as a
function of the charge of the nanotube has the similar behavior. 
\begin{figure}
\includegraphics[width=0.4\textwidth, height=0.2\textheight]{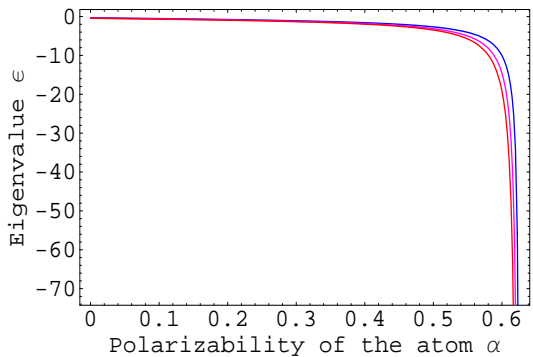}
\caption {(color online) A plot of binding energy of atom as a
function of the polarizability
 $\alpha$ of the atom and with fixed line charge $q=1$ of the charged wire and
 for three different values of the self-adjoint extension parameter
 $\omega$. From top to bottom $\omega= \frac{\pi}{6},
 \frac{\pi}{8},\frac{\pi}{10}$ respectively.}
\end{figure}

We now address the important issue of smearing observed \cite{try}
in quantized  conductance(QC). From our analysis of bound state
formation we see that for $\eta^2\in[0,1)$, there is a single bound
state, which depends on the self-adjoint extension parameter
$\omega$. By suitably adjusting the parameter $\omega$, this  bound
state eigenvalue  can be made large negative value.  Now these
deeply bound states will then contribute to the smearing of the edge
of quantized conductance.

Let us now concentrate on breaking of scale invariance because of
nontrivial quantization of the system. Since scale invariance is
broken upon quantization, it naturally termed  as quantum mechanical
anomaly. Before discussing scale symmetry breaking let us discuss
scale symmetry of this system by looking at its classical version.
One can construct classical Lagrangian from the Hamiltonian
(\ref{radial}) as $L= \frac{d^2r}{dt^2}-\frac{1}{r^2}\left(\eta
^2-\frac{1}{4}\right)$. This lagrangian scales as
$\frac{1}{\lambda^2}L$ under the scale transformation $r\to\lambda
r$ and $t\to\lambda^2 t$. One can also see  that the action
$\mathcal A=\int dtL$ for the system is invariant under the scale
transformation. So the system is classically scale invariant. Scale
invariance implies that if $\chi$ is a eigenstate of the Hamiltonian
(\ref{radial}) with eigenvalue $-E,~~ E>0$, then
$\chi_\lambda=\chi(\lambda r)$ will also become an eigenstate of
the same Hamiltonian (\ref{radial}) with eigenvalue $-E/\lambda^2$.
By tuning $\lambda$ one can get any negative eigenvalue up to
$-\infty$. It indicates that the system must not have any bound
state. But we see that after quantization the system admits bound
states for specific values of the parameter $\eta$. This violation
of scale symmetry due to nontrivial quantization can be understood
as follows. We first discuss the case for $\eta \neq 0$. In order to
discuss scaling anomaly quantum mechanically we define  a scaling
operator $\Lambda = \frac{-i}{2} (r \frac{d}{dr} + \frac{d}{dr} r)$.
It acts on any  element $\phi$ belonging to the domain
$D_\omega(H_r)$. In our case this scaling operator $\Lambda$ takes
the wave-function $\phi$ out of the domain $D_\omega(H_r)$. It can
be easily understood from the relation
\begin{eqnarray}
\nonumber \lim_{r\to 0}\Lambda \phi(r) \simeq \hspace{5cm}\\
\nonumber \frac{(1+\eta)}{i} \frac{2e^{i\omega/2}}{\Gamma(1+\eta)}
\frac{\sin(\omega/2+3\pi\eta/4)}{\sin\pi\eta}\left(\frac{r}{2}\right)^{\eta+1/2}\\
-
\frac{(1-\eta)}{i}\frac{2e^{i\omega/2}}{\Gamma(1-\eta)}\frac{\sin(\omega/2+
\pi\eta/4)}{\sin\pi\eta}\left(\frac{r}{2}\right)^{-\eta+1/2}
\label{scale1}
\end{eqnarray}
Because of the two different multiplying  factors $(1 + \eta )$ and
$(1 - \eta )$ in (\ref{scale1}), $\Lambda \phi(r) \neq  C \phi(r)$
($C$ is any complex constant) . So clearly $\Lambda\phi(r)$ does not
belong to the domain $D_\omega(H_r)$.  This shows that scale
invariance is thus broken after inequivalent quantization of the
system by self-adjoint extensions. However, scaling anomaly is not present
for all values of the self-adjoint extension parameter $\omega$. For
example, for $\omega = -\frac{ \eta \pi}{2}$ and $ \omega = -\frac{
3 \eta \pi}{2}$, the action of the scaling operator $\Lambda$ on any
function $\phi(r) \in D_\omega(H_r)$ does not throw it outside the
domain. For these two values of the parameter $\omega$  the scaling
symmetry is restored \cite{kumar1,kumar2,kumar3} and there does not
exist  any bound solution. We will not discuss $\eta=0$ case here,
but one can discuss it similarly. For negative values of the
parameter $\eta^2$, the usual analysis gives rise to bound states
whose ground state goes to $-\infty$. It is however possible  to use 
re-normalization group
techniques \cite{rajeev}. But since it is in strong attractive
region, it does not have so much importance as far as semaring is
concerned.

In conclusion, here we  shown that it is possible for polarizable
neutral atom to form a single bound state in the electric field of
the charged nanotube (SWNT). Nontrivial boundary conditions obtained
by self-adjoint extensions makes it possible to get this bound state
solution. We have used von Neumann method to make the initially non
self-adjoint Hamiltonian $H_r$ self-adjoint. We have found out the
domain $D_\omega(H_r)$ on which the Hamiltonian is self-adjoint. We
have shown that in general the scaling operator $\Lambda$ acting on
this domain $D_\omega(H_r)$ does not keep the domain invariant. But
for two values of the self-adjoint extension parameter $\omega =
-\frac{ \eta \pi}{2}$ and $ \omega = -\frac{ 3 \eta \pi}{2}$,
scaling symmetry does not brake down. The existence  of a bound
state may contribute to the possible smearing of the edge of
quantized conductance. The detection of atomic binding in the the
electric field of the charged nanotube would be an interesting
experiment. At least, within the experimental setup in  Ref. \cite{try}
one should see the effect of bound state in scattering or absorption
cross section.

\vskip 0.5 cm

\noindent {\bf Acknowledgment}\\ We thank Kumar S. Gupta for
encouraging us and helping us all the time throughout the
preparation of the manuscript. The careful reading of the manuscript
by P. B. Pal is also acknowledged.

\end{document}